\documentclass{PoS}

\newcommand{\point}{\quad .}
\newcommand{\comma}{\quad ,}
\newcommand{\D}{\mathcal{D}}

\newcommand{\Aslash}{\ensuremath \raisebox{0.025cm}{\big\slash}\hspace{-0.25cm} A}
\newcommand{\dslash}{\not{\hbox{\kern-2pt $\partial$}}}

\title{Baryon currents in the C-broken phase of QCD}

\ShortTitle{Baryon currents}

\author{Biagio Lucini\\
        Physics Department, Swansea University, Singleton Park, Swansea SA2 8PP, UK\\
        E-mail: \email{b.lucini@swansea.ac.uk}}

\author{\speaker{Agostino Patella}\\
        Scuola Normale Superiore, Piazza dei Cavalieri 27, 56126 Pisa, Italy\\
        and INFN Pisa, Largo B. Pontecorvo 3 Ed.~C, 56127 Pisa, Italy\\
        E-mail: \email{patella@sns.it}}

\author{Claudio Pica\\
        Physics Department, Brookhaven National Laboratory, Upton, NY 11973, USA\\
        E-mail: \email{pica@bnl.gov}}

\abstract{In a space with some sufficiently small compact dimension (with non-trivial cycles) and with periodic boundary conditions for the fermions, the charge conjugation (C), spatial parity (P), time reversal (T) and CPT symmetries are spontaneously broken in QCD. We have investigated what are the physical consequences of the breaking of these discrete symmetries, that is what local observables can be used to detect it. We show that the breaking induces the generation of baryon currents, propagating along the compact dimensions.}

\FullConference{The XXV International Symposium on Lattice Field Theory\\
		 July 30-4 August 2007\\
		 Regensburg, Germany}

\begin{document}

\section{Introduction}

It is well known~\cite{vanBaal:1988va} that, in a space with some sufficiently small compact dimension (with non-trivial cycles) and with periodic boundary conditions for the fermions, the charge conjugation (C), spatial parity (P), time reversal (T) and CPT symmetries are spontaneously broken in QCD. Recently, some work was done on this subject, following studies about the orientifold planar equivalence.

The orientifold planar equivalence~\cite{Armoni:2003gp} is the equivalence in the planar limit of two theories: QCD with one massless quark in the antisymmetric representation on one side; and the $\mathcal{N}=1$ super Yang-Mills, on the other side. The equivalence was proved firstly in~\cite{Armoni:2004ub}; in this proof it was implicitly used that the C-symmetry is not spontaneously broken. An alternative proof can be found in~\cite{Unsal:2006pj}; it has the merit of clarifying that the invariance of the vacuum under C is also a necessary condition for the validity of the orientifold planar equivalence. Both the dependence of the symmetry breaking on the boundary conditions, and the breaking of CPT give convincing evidence that all the discrete symmetries are restored at a large enough size of the compact dimensions.

In our work~\cite{Lucini:2007im}, we have investigated what are the physical consequences of the breaking of the discrete symmetries, that is what local observables can be used to detect it. Since CP (as well as PT and CT) remains unbroken, we can easily say what observables cannot be used: the masses of particles and antiparticles are still degenerate, the baryon number is still conserved. We will show that the breaking induces the generation of baryon currents, propagating along the compact dimensions. It is worth noticing that a non-zero expectation value for a current breaks the Lorentz symmetry, and therefore it must be zero in the limit of infinite volume.

\section{Breaking of discrete symmetries}

We consider a non-Abelian gauge theory with gauge group SU($N$) and $N_f$ families of fundamental Dirac fermions. The manifold on which the theory is defined is $\mathbf{R}^d \times T_n$, where $d+n=4$ and $T_n$ is a spatial $n$-dimensional torus. We assume that the torus has the same extension $R$ in all the compact dimensions. Moreover, spatial directions are closed with periodic boundary conditions for both fermions and bosons.

In what follows, $(x_a)_{a=1, \dots, d}$ are the coordinates on $\mathbf{R}^d$, while $(z_\alpha)_{\alpha=1, \dots, n}$ are the coordinates of the compact dimensions.

We compute the vacuum expectation value ({\em vev}) of the Wilson line in a compact direction~\cite{Unsal:2006pj}
\begin{equation}
W^{(A)}_{\alpha} (x, z) = \mathrm{Pexp} \; \left( i \int_0^{R} A_\alpha (x,z) \, d z_\alpha \right) \point
\end{equation}
The gauge can be fixed in such a way that $W$ is diagonal 
\begin{equation}
W^{(A)}_\alpha (x, z) = \left( e^{iv^{(A)}_{1 \alpha}(x, z)}, \dots , e^{iv^{(A)}_{N \alpha}(x, z)} \right) \point
\end{equation}
We focus on the effective potential for	those eigenvalues\footnote{ We drop the subscript $\alpha$ where this does not lead to ambiguities.}:
\begin{eqnarray}
e^{i V_d V(v_1, \dots , v_N)} &=&
\int \, e^{ i S } \, \prod_k \delta \left( v_k - \frac{1}{V_d R^n} \int v^{(A)}_k(\mathbf{x}) \, dx dz \right) \, \D A \D \bar{\psi} \D \psi \point
\end{eqnarray}
The absolute minima of $V(v_1, \dots , v_N)$ give the expectation values for the set of the eigenvalues. If C-symmetry is broken, the set of eigenvalues is not invariant under the substitution $v \rightarrow -v$. In the one-loop approximation, the effective potential for the Wilson line is given by~\cite{Barbon:2006us}
\begin{eqnarray}
\label{onelooppotential}
&& V(\vec{v}_1, \dots, \vec{v}_N) = \left[ \sum_{i,j = 1}^{N} f(0, \vec{v}_i-\vec{v}_j) - 2 N_f \sum_{i=1}^{N} f(m, \vec{v}_i) \right] \comma\\
&& \textrm{with } f(m,\vec{v}) =  \frac{1}{R} \left( \frac{m R}{\pi}\right)^2 \sum_{\vec{k} \neq 0} \frac{K_2( m R k)}{k^2} \sin^2 \left( \frac{1}{2} \vec{k}\cdot \vec{v} \right)
 \comma
\end{eqnarray}
where $K_2$ is the modified Bessel function of the second kind of order 2 and the sum runs over $n$-index integer vectors $\vec{k}$. The first term of eq.~(\ref{onelooppotential}) gives rise to an attraction between the eigenvalues. The second term produces an unconstrained absolute minimum at $v_{i} = \pi$. When the $SU(N)$ constraint is taken into account, the minima are:
\begin{equation}
v_{1}^* = v_{2}^* = \dots = v_{N}^* = \left\{
\begin{array}{ll}
\pm \frac {N-1}{N} \pi & \quad \textrm{for $N$ odd}\\
\pi \qquad & \quad \textrm{for $N$ even}\\
\end{array}
\right. \point
\end{equation}

The spontaneous symmetry breaking shows up as a non-zero imaginary part of the Wilson line. Hence, if $N$ is even, there is no symmetry breaking. If $N$ is odd, then P, C, T, CPT are broken. There are $2^n$ vacua. This result is due to the spacial periodic boundary conditions for the fermions. The validity of the calculation in the non-perturbative regime has been checked on the lattice in~\cite{DeGrand:2006uy}, and the discrete symmetries are shown to be broken for $R$ below the fermi scale.

\section{Baryon currents}

The Wilson line is an order parameter for the symmetry breaking. However, if the theory is not invariant under P, C, T, we expect this to be reflected by some local observable. This observable must be odd under the broken symmetries, but even under the combined action of two of them. The spatial components of the baryon current $j_\alpha = \langle \sum_{i=1}^{N_f} \bar{\psi}_i \gamma_\alpha \psi_i \rangle$ satisfy this property.

Why should we expect a non-zero baryon current? A non trivial {\em vev} of the Wilson line means a non-zero value of the gauge field in that direction. Since the system is translationally invariant along the compact direction, the value of the gauge field must be constant (note that in the presence of toroidal topology a constant field cannot be gauged away). The background gauge field acts as a non-trivial source for the baryon current. Hence, we expect this current to be different from zero.

In order to compute the expectation value of the baryonic current, we define the partition function of the system in presence of a generalised "chemical potential" $\mu_\alpha$:
\begin{equation} \label{partitionfunction}
Z(\mu) =
\int \, \exp \left\{ i S_G + i \sum_{f=1}^{N_f} \int \bar{\psi}_f(x) \left( i \dslash - \Aslash - \mu_\alpha \gamma_\alpha - m \right) \psi_f(x) \, d^4 x \right\} \, \D A \D \bar{\psi} \D \psi \point
\end{equation}
Since the insertion of the $\mu$-term has the same effect as shifting $A_\alpha(\mathbf{x},x_\alpha) \rightarrow A_\alpha(\mathbf{x},x_\alpha) + \mu_\alpha \mathbf{1}_N$ (the gauge action is not affected by this shift), or as shifting the phases of the eigenvalues of the Wilson lines $v_{\alpha k}(\mathbf{x}) \rightarrow v_{\alpha k}(\mathbf{x}) + R \mu_\alpha$, the partition function can be obtained as:
\begin{equation}
Z(\mu) = e^{i V_d V(v_1^* + R\mu, \dots , v_N^* + R\mu)}
\end{equation}
On the other side, as the chemical potential in~\ref{partitionfunction} acts like a source for the current, which is obtained by deriving the partition function with respect to $\mu_\alpha$:
\begin{equation}
j_\alpha = \frac{i}{V_d R^n} \frac{d \, \log Z}{d \mu_\alpha}(0) = - \frac{1}{R^{n-1}} \sum_k \frac{\partial V}{\partial v_k}(v_1^*, \dots , v_N^*) \point
\end{equation}

\begin{figure}
\begin{center}
\includegraphics[width=1\textwidth]{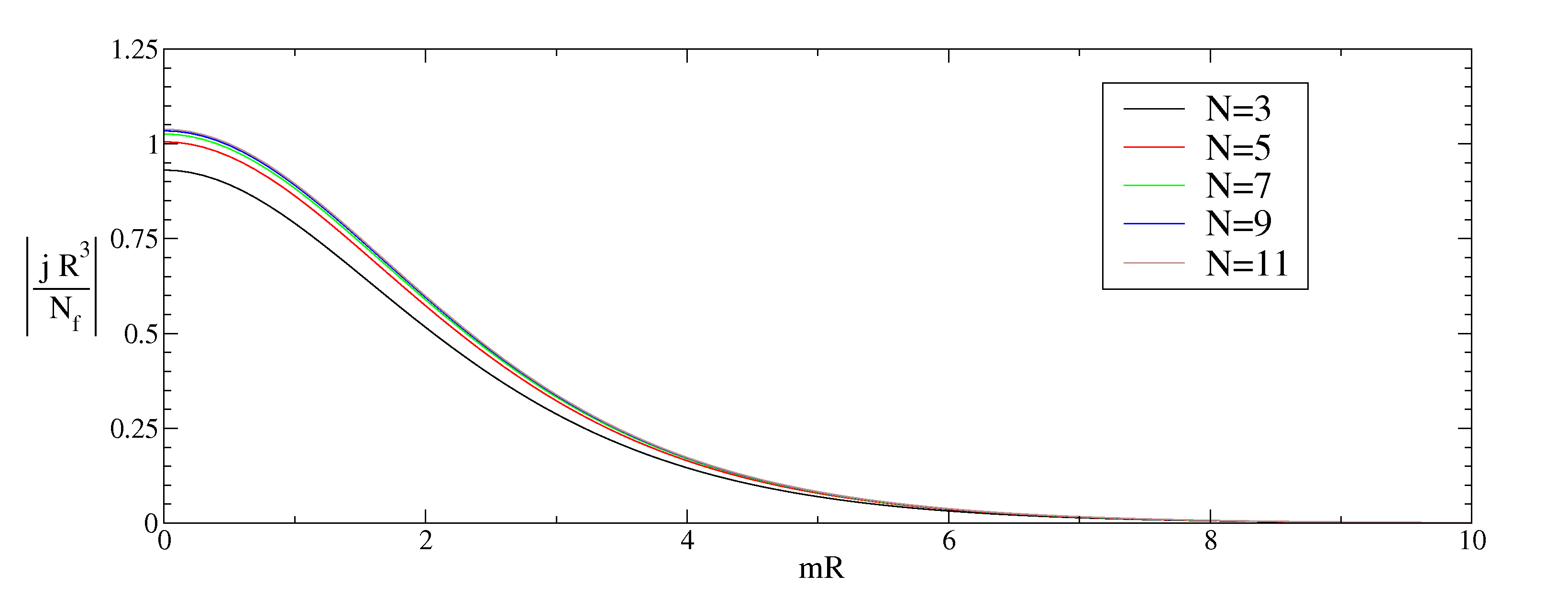}
\end{center}
\caption{The baryon current (one-loop approximation) vs. $mR$ at different values of $N$.}
\label{current}
\end{figure}

Using the one-loop expression for the effective potential in~\ref{onelooppotential}, we obtain for the baryon current:
\begin{eqnarray}
\langle j_\alpha \rangle = - \frac{N_f N (mR)^2}{R^3 \pi^2}
\sum_{\vec{k} \neq 0} \frac{ K_2(mkR) k_\alpha \sin ( \vec{k}\vec{v}^* ) }{k^2} \point
\end{eqnarray}
$\langle j_{\alpha} \rangle$ is zero when $v_{\alpha}^* = 0$ or  $v_{\alpha}^* = \pi$ (i.e. when the symmetry breaking does not occur in direction $\alpha$), is odd under $v_{\alpha}^* \to -v_{\alpha}^*$, goes to zero when $m \to \infty$ (fig.~\ref{current}).

\section{The lattice calculation}

\begin{figure}
\begin{center}
\includegraphics[width=0.6\textwidth]{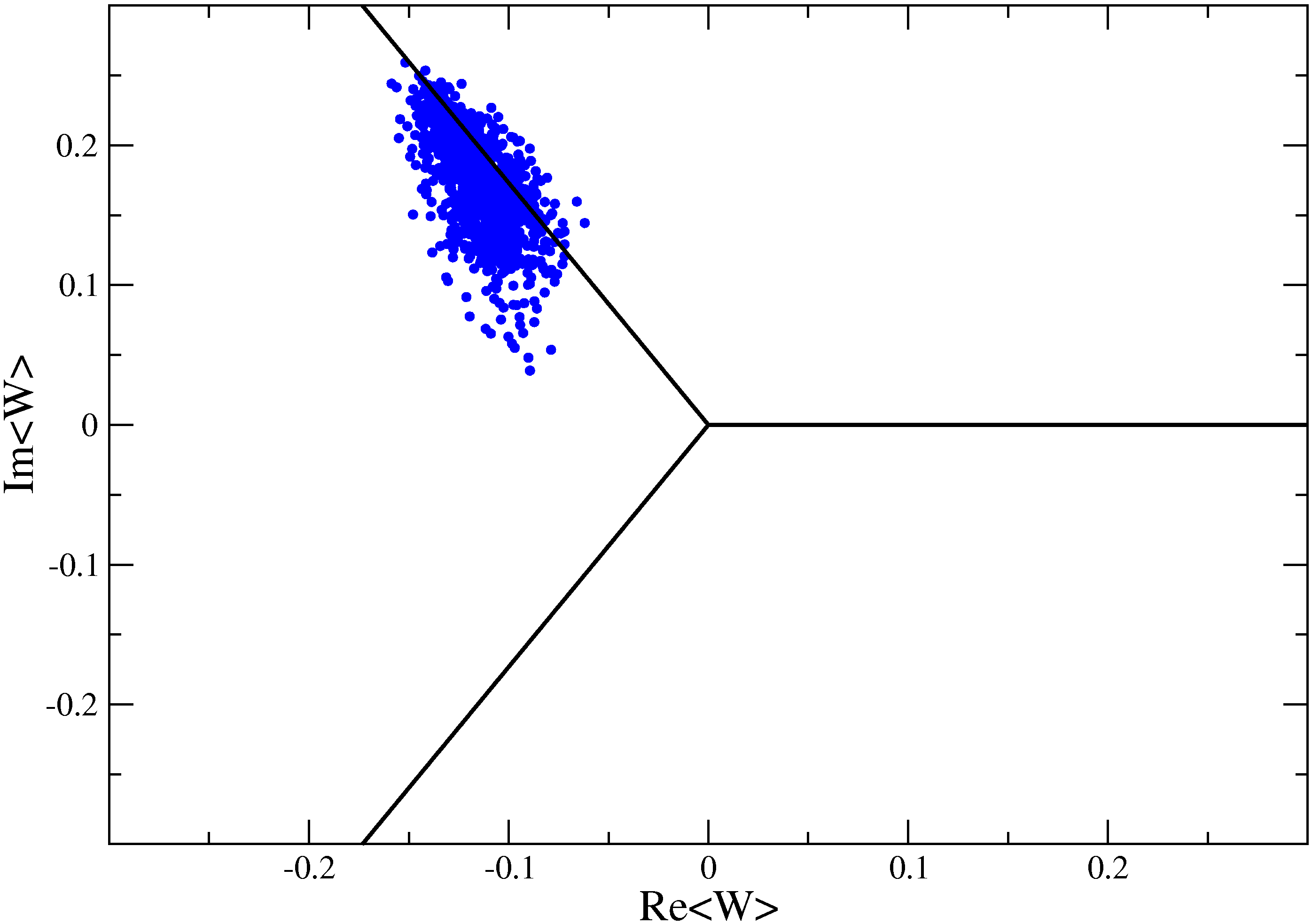}
\end{center}
\caption{Plot of the Wilson loop. Clustering of the phase around $2 \pi/3$ means spontaneous breaking of discrete symmetries.}
\label{wloopplot}
\end{figure}

Details of the lattice action used in our simulation are provided in~\cite{Lucini:2007im}. A $24 \times 4^3$ lattice at $\beta = 5.5$ was used. This fixes the lattice spacing (determined measuring the Sommer parameter $r_0$~\cite{Sommer:1993ce}) to 0.125 fm. This means that $L_t  = a N_t = 3$ fm and $L_s = a N_s = 0.5$ fm, in agreement with the assumptions of sufficiently large $L_t$ and $L_s$ below the fermi scale. The breaking of discrete symmetries can be checked by looking at the Wilson line wrapping around a spatial direction. A typical behaviour is plotted in fig.~\ref{wloopplot}, which shows that the Wilson loop magnetises along $e^{i \frac{2}{3} \pi}$, i.e. $\langle W \rangle$ acquires an imaginary part, as required by the symmetry breaking scenario.

\begin{figure}
\begin{center}
\includegraphics[width=0.6\textwidth]{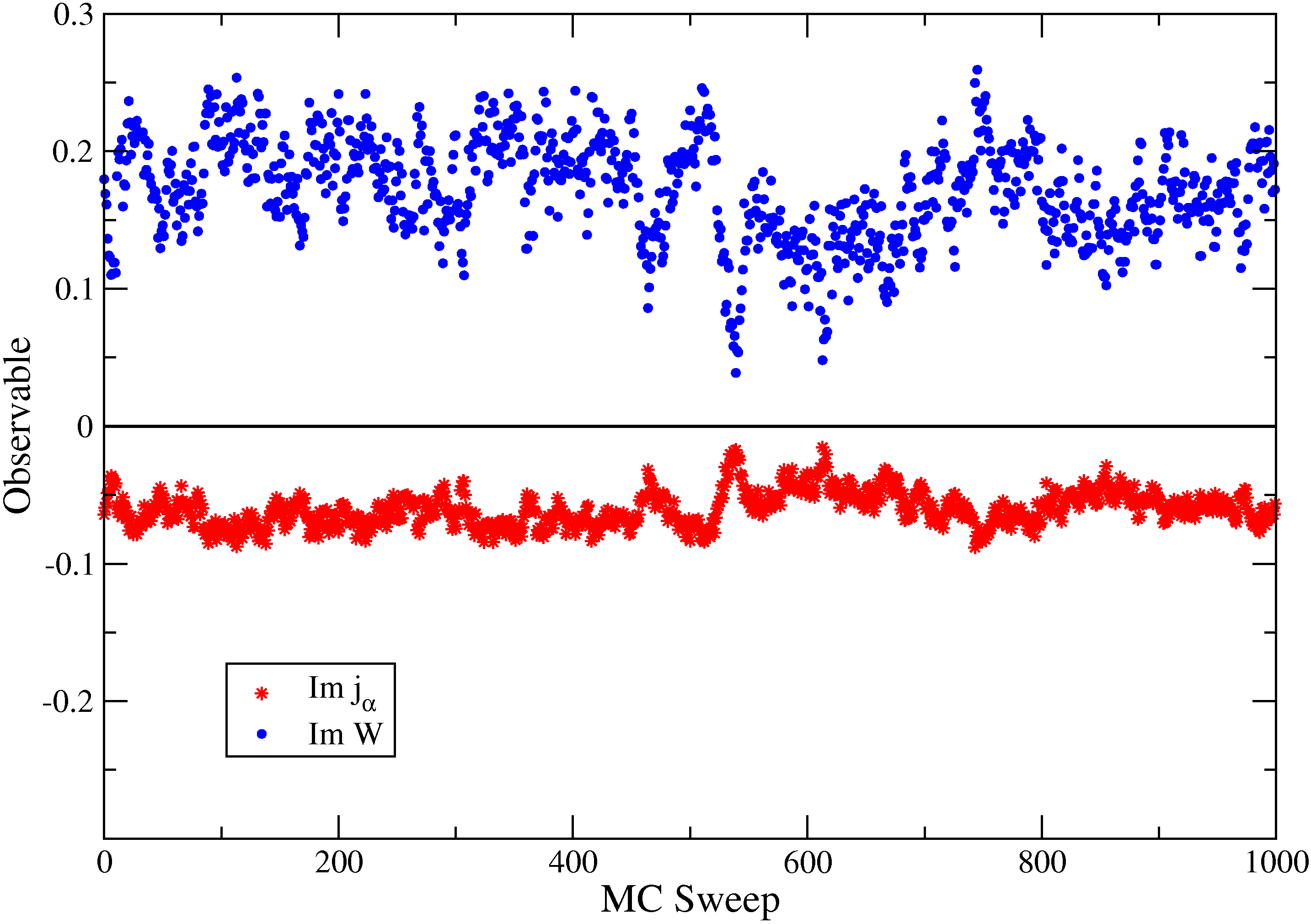}
\end{center}
\caption{Monte Carlo history of the imaginary part of the Wilson line and the current in a compact direction in the broken symmetry phase.}
\label{figcurrent}
\end{figure}

For the Euclidean rotated theory in the broken phase, we expect a non-zero value of the imaginary part of the current. At the given lattice parameters we find $|\mbox{Im}\langle j_{\alpha} \rangle| = 0.060 \pm 0.002 $ (fig.~\ref{figcurrent}), which should be compared with the perturbative prediction $\langle j_{\alpha} \rangle \simeq 0.037473(4)$: it is remarkable that for compact dimensions of the order of $1/\Lambda_{QCD}$ the perturbative prediction still gives the correct order of magnitude.

\begin{figure}
\begin{center}
\includegraphics[width=0.6\textwidth]{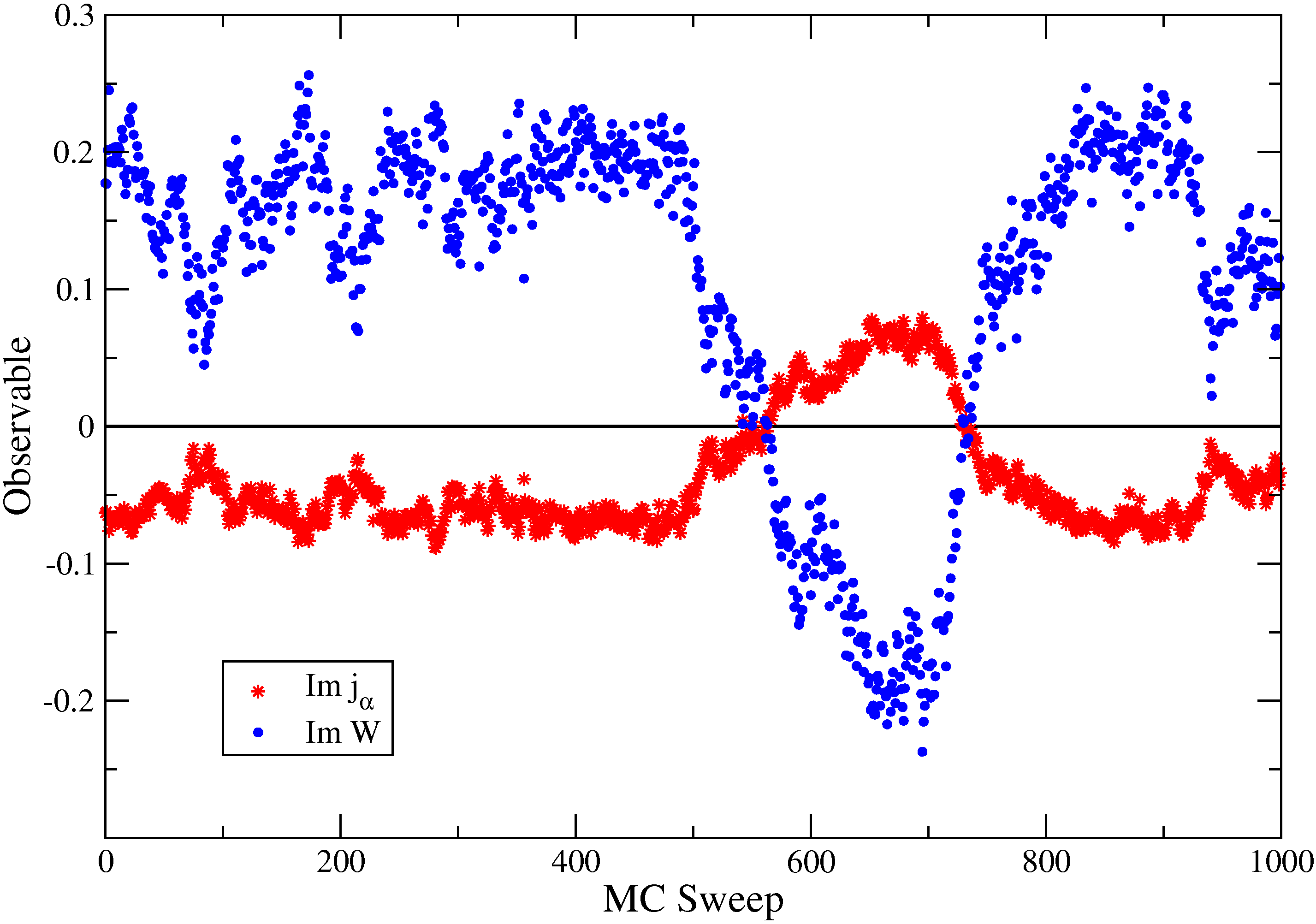}
\end{center}
\caption{Tunneling, due to the finite volume effects.}
\label{figtunneling}
\end{figure}

Since in simulations all the dimensions are kept finite, tunneling among the vacua are expected. In fig.~\ref{figtunneling} two consecutive tunneling events are shown. The correlation of the current with imaginary part of the Wilson line is evident: both change their sign when tunneling occurs.

\section{Conclusions and outlook}

In the phase in which discrete symmetries are spontaneously broken, there is a persistent baryonic current wrapping around the topologically non-trivial compact directions. This current is similar to the supercurrent observed in superconductors. However, there is a fundamental difference: unlike the case of superconductors, in QCD in compact not simply connected space the current is still conserved, since the U(1) baryon symmetry (which in the case of QCD is a global symmetry) remains unbroken. The persistent flow is induced by the spontaneous breaking of discrete symmetries.

In this case, pairs of quarks and antiquarks condense in the vacuum, while the total baryonic number is zero. The quark and the antiquark move with opposite momenta along the compact dimensions. The total momentum is zero, but there is a net baryonic current.

The existence of this current is a clear physical signature of the symmetry breaking and could be used to determine the order of the symmetry restoring phase transition that happens at a critical radius of the compact direction, which string-inspired calculations predict to be second order~\cite{Armoni:2007jt}. Another open problem is the interplay between the aforementioned phase transition and the chiral symmetry restoring phase transition at finite temperature.

\end{document}